\documentstyle [twocolumn]{mn}
\newcommand{\mincir}{\raise
-2.truept\hbox{\rlap{\hbox{$\sim$}}\raise5.truept\hbox{$<$}\ }}
\newcommand{\magcir}{\raise
-2.truept\hbox{\rlap{\hbox{$\sim$}}\raise5.truept\hbox{$>$}\ }}
\newcommand{\minmag}{\raise
-2.truept\hbox{\rlap{\hbox{$<$}}\raise6.truept\hbox{$<$}\ }}
\newcommand{\be}{\begin{equation}}
\newcommand{\ee}{\end{equation}}
\newcommand{\ba}{\begin{eqnarray}}
\newcommand{\ea}{\end{eqnarray}}
\newcommand{\brr}{\begin{array}}
\newcommand{\err}{\end{array}}
\newcommand{\bc}{\begin{center}}
\newcommand{\ec}{\end{center}}

\title[Modelling the X--ray cluster dipole and cluster contribution to
the soft X--ray background]
       {Modelling the X--ray cluster dipole and cluster contribution to
the soft X--ray background} 
\author[V. Kolokotronis et al. ]
{V. Kolokotronis$^{1}$, M. Plionis$^{2}$, P. Coles$^{1}$ and 
S. Borgani$^{3,4}$\\
$^1$ Astronomy Unit, School of Mathematical Sciences,
Queen Mary and Westfield College, Mile End Road, London E1 4NS \\
$^2 $ National Observatory of Athens, Lofos Nimfon, Thesio, 18110 Athens,
Greece \\
$^3$ INFN Sezione di Perugia, c/o Dipartimento di Fisica dell' Universit\`{a}, 
via A. Pascoli, I-06100 Perugia, Italy \\
$^4$ SISSA - International School for Advanced Studies,
via Beirut 2-4, I-34013 Trieste, Italy \\}

\begin{document}
\maketitle

\begin {abstract}
We investigate the 
sampling and dipole convergence properties
of flux--limited samples of mock X--ray clusters in relation to their
underlying ``parent'' cluster distribution. To this purpose, we resort
to numerical simulations of the cluster distribution and extract
samples resembling the main observational features of X--ray selected
cluster samples. The flux--limited samples, being quite sparse,
underestimate the amplitude of the ``parent'' cluster dipole by
$\approx$ 15 per cent on average for Local Group-like
observers. However, the general shape of their dipole amplitude
profiles are in relatively good agreement. We also calculate the
expected contribution of clusters, selected according to the relevant
criteria, to the soft (i.e. $0.1-2.4$ keV) extragalactic X--ray
background, using the ESO Key Project X--ray luminosity function,
assuming a flat universe with vanishing cosmological constant. We
obtain a value of about 10 per cent of the observed XRB flux. 

\vspace{5mm}

\end{abstract}

\vspace{5mm}
\section{Introduction}
Clusters of galaxies are the largest
gravitationally--collapsed structures in the universe. This property
makes these objects potentially useful tracers of the global matter
distribution. Considerable observational effort has thus been spent in
compiling cluster catalogues as free as possible from 
errors and selection biases.  While it remains a very
difficult task to construct bias--free  optically--selected samples of 
rich clusters, selecting objects through their X--ray emission offers
a cleaner possibility (e.g. Cen 1997). 
The powerful X--ray emission from the hot intra-cluster 
gas, peaked at the gravitational centre of the clusters potential wells, 
defines them much better than optical light and it is more
directly related to their total mass, making such X--ray emitting objects prime
candidates for tracers of the matter distribution on very large
scales (for recent reviews see Ebeling et al. 1995;
Guzzo et al. 1995; Collins et al. 1995; Gioia 1996).

The analysis we shall undertake in this paper was initially
inspired by the {\small ROSAT} all-sky survey (RASS) and the
follow-up work forming the ESO Key Program on southern cluster
redshifts (Guzzo et al. 1995; de Grandi 1996). A number of 
other X--ray cluster surveys based on {\small ROSAT} are also available; 
the South Galactic Pole 
project (SGP) by Romer et al. (1994) (a flux--limited sample comprising
128 X--ray clusters restricted to an area of 3100 ${\rm deg}^{2}$), the
extensive optical follow-up work aimed at selected clusters from RASS 
by Allen et al. (1992) and by Crawford et al. (1995) (that accompanied
and led to the X--ray compilation of the brand new Brightest Cluster
Sample (BCS) by Ebeling et al. 1997b) and last but not least at all,
the recent X--ray Brightest Abell-type cluster sample (XBACs), which
constitute the largest all-sky flux--limited cluster sample to date
(Ebeling et al. 1996b).
Prior to the {\small ROSAT} mission many other X--ray projects were
conducted with the aim to construct a large unbiased X--ray
flux--limited cluster sample that could be used for detailed cosmological and
statistical purposes (mapping the large scale structure of the
universe, quantifying cluster evolution, investigating correlation
function properties, establishing the X--ray luminosity function for
clusters of galaxies, to name but a few).

The principal  aim of this work is to investigate the disparities of dipole
amplitude, structure and alignment between the X--ray flux--limited samples and
their ``parent'' cluster distribution. 
We shall investigate this using `mock' cluster
samples  generated in numerical simulations via an optimized version of the 
Truncated Zel'dovich approximation (TZA; Coles, Melott \& Shandarin 1993;
Borgani, Coles \& Moscardini 1994; Borgani et al. 1995). 
This study therefore complements and extends that of
Kolokotronis et al. (1996) and of Tini Brunozzi et al. (1995) (TB95
hereafter) which concerned the relationship between the dipole of
optically-identified clusters, that of galaxies, and that of the
underlying matter distribution in simulations.
The authors found that the high peaks of the density
field (clusters) do trace the underlying mass dipole structure
reasonably well, with a roughly constant amplitude difference
corresponding to a linear biasing term, though with a relatively large
observer-to-observer scatter.
The degree of cluster-mass dipole correlation depends on the density
threshold above which the clusters are defined and for the relatively
low density threshold, which corresponds to a mean separation of $\sim
30h^{-1}\,$Mpc (similar to that of the ESO KP X-ray clusters), they
found a very good correlation. However, the above cluster-mass dipole
correlation starts degrading with increasing intercluster separation,
or density threshold (see section 4.1 of Kolokotronis et al. 1996). 

Dipole analyses have been carried out by many authors in recent years; 
for a complete description of the literature, see the references in 
Kolokotronis et al. (1996) (hereafter K96).

Using linear perturbation theory (Peebles 1980) one can relate the cluster 
dipole and the observer peculiar velocity to obtain an estimate of the 
$\beta_{\rm c}$-parameter ($\equiv \Omega_0^{0.6}/b_{\rm c}$, where 
$b_{\rm c}$ is the cluster bias parameter with respect to the mass).
It is worth noting here specifically that, different classes of
extragalactic objects  (QSOs, AGN, galaxies, clusters of galaxies) 
have different relationships to the underlying matter distribution
and may therefore differ substantially in their respective dipole
behavior. Although this complicates the use of dipole properties
as estimators of the cosmological density parameter, $\Omega_0$,
it does allow one in principle to study the relative bias displayed
by such objects (K96; Plionis 1995). 
Prominent among those analyses that have used X--ray selected objects 
as tracers of the density field are Miyaji $\&$ Boldt (1990) (X--ray AGN) 
and Lahav et al. (1989) (X--ray clusters).

Our subsidiary intention, having generated the mock X--ray cluster samples
complete down to a certain flux limit, is to estimate their contribution to 
the soft X--ray Background (XRB) which, ever since its discovery 
(Giacconi et al. 1962), has been one of the outstanding puzzles in this 
field. 
Such an application is particularly interesting because as far as we know we 
will be simulating a large sample of X--ray clusters 
(for similar X--ray samples see also McKee et al. 1980; 
Piccinotti et al. 1982; Maccacaro et al. 1984; Lahav et al. 1989; 
Gioia et al. 1990; Edge, Stewart, Fabian \& Arnaud 1990; Edge \&
Stewart 1991; Henry et al. 1992; Guzzo et al. 1995; de Grandi 1996;
Ebeling et al. 1995; Ebeling et al. 1997b and references therein).  
This enables us to place tight limits on the relation between
X--ray emitting clusters and the XRB in the soft energy band (0.1-2.4 keV).
We also stress the fact that, a substantial number of work is dedicated
to measure the contribution of X--ray objects to the hard XRB
(2-10 keV). Although the availability of data from the
{\small ROSAT} satellite makes this task feasible for
energies below 2 keV, the fact that both the
spectrum of the XRB and the contribution of the resolved discrete
component are yet to be determined, forces us to make some relatively 
uncertain assumptions, as we point out later.

The outline of the paper is as follows.  An overview of the dipole
analysis, together with techniques for an accurate determination of the 
inherent shot-noise errors, is presented in
Section 2. We present a useful set of definitions, 
observational constraints and the method of reproducing
reliable simulations of the relevant X--ray
populations in section 3. 
We discuss the cluster dipole results in section 4, while an estimate
of the contribution of X--ray clusters to the soft XRB in a critical
density model is presented in section 5. Finally, we present our main
conclusions in section 6. 

\section{Dipole Formalism}
In this section we will present the method we use to calculate the
peculiar gravitational acceleration induced by the distribution of
matter as it is traced by clusters of galaxies. 

\subsection{Theoretical Preamble}
As we discussed more extensively in K96,
we need to assume the standard gravitational instability picture
for all of this work. Within this picture we assume that linear
perturbation theory applies and, perhaps with  less justification, that
a linear biasing scheme relates fluctuations in cluster numbers to
fluctuations in the overall density of material, i.e.
\begin{equation}
\left(\frac{\delta \rho}{\rho}\right)_{\rm clusters} = b_{\rm c} 
\left(\frac{\delta \rho}{\rho}\right)_{\rm mass}
\end{equation}
(Kaiser 1984; Coles 1993). This scheme motivates the introduction of the
parameter $\beta$, defined by
\begin{equation}
\beta \equiv \frac{f(\Omega_{\circ})}{b} \simeq 
\frac{\Omega_{\circ}^{0.6}}{b},
\end{equation}
where $f$ is the mass fluctuation growth rate in linear perturbation theory 
(e.g. Peebles 1980). 
Using linear theory we can relate velocities and accelerations. A classical 
case is to relate the only very accurately known
observed peculiar velocity in cosmology, that of the Local group (LG) 
relative to the cosmic microwave background (CMB), ${\mbox{\boldmath$u(r)$}}$, 
to the peculiar gravitational acceleration induced upon it
by the surrounding density fluctuations, ${\mbox{\boldmath$g(r)$}}$,
which then, provided that the two vectors are aligned, allows $\beta$
to be estimated via:
\begin{equation}\label{eq:lpt}
{\mbox{\boldmath$u(r)$}}=\frac{2}{3}
\frac{f(\Omega_{\circ}){\mbox{\boldmath$g(r)$}} }
{H_{\circ} \Omega_{\circ}}=\frac{\beta}{4 \pi} \int \delta({\mbox{\boldmath$r$}}) 
\frac{{\mbox{\boldmath$r$}}}{|{\mbox{\boldmath$r$}}|^3} {\rm d}r,
\end{equation}
where $\delta({\mbox{\boldmath$r$}})$ is the density fluctuations of
the underlying mass distribution at position ${\mbox{\boldmath$r$}}$.
Knowledge of
$b$ allows an estimate of $\beta$, if $\Omega_{\circ}$ is given as an
input parameter and vice versa. Although the assumption (1) is relatively
well--motivated in a statistical sense, it may well turn out to be a poor 
approximation when applied on a point-to-point basis, as required by this 
analysis. In fact, K96 found by comparing mass and cluster dipoles in 
simulations, in which $\Omega_{\circ}$ was an input parameter,
that they recovered the correct mean bias value but albeit with a large
scatter (see also section 4.4).

Note that we will be using distance units in ${\rm km} \; {\rm s}^{-1}$ and 
thus the $H_{\circ}$ dependence drops. Wherever necessary, we will use 
$H_{\circ}=100h\,{\rm km}\,{\rm s}^{-1}\,{\rm Mpc}^{-1}$.

\subsection{Analysis Tools}
We employ the method of moments to quantify the distribution of
clusters around the observer.
Having an available set of cluster positions,
${\mbox{\boldmath$r$}}_{\rm{i}}$, around an observer placed at the origin,
we can calculate the dipole and  monopole moments as follows:
\begin{equation}\label{eq:mon}
M=\sum_{{\rm {i}}=1}^{N(R)} w_{\rm{i}}
\end{equation}
and
\begin{equation}\label{eq:dip}
{\mbox{\boldmath$D$}} =\sum_{\rm{i}=1}^{N(R)} w_{\rm{i}} \hat{{\mbox
{\boldmath$r$}}}_{\rm i}.
\end{equation}
The sum is taken over all $N(R)$ clusters within a radius $R$
and $w_{\rm i}\,(\propto r_{\rm i}^{-2})$ are
the appropriate weights, which should also take into account any sample
selection effects. The monopole should increase linearly with distance $R$ 
while the dipole ${\mbox{\boldmath$D$}}$ keeps increasing until the
most distant density inhomogeneity, that affects the dynamics of the
observer, is taken into account. The scale at which this happens, which 
we denote as the convergence depth ($R_{\rm conv}$), should be
within the effective depth of the catalogue if the estimated value of 
$\beta$, via equation (\ref{eq:lpt}) is to be reliable.

Since the samples we are interested in are flux--limited,
the number of objects decreases with
radial distance due to the rapidly decreasing selection function
$\phi(r)$, defined as the number density of objects that can be seen above 
the limiting luminosity at distance $r$:
\begin{equation}\label{eq:self}
\phi(r)=\int_{L_{{\rm min}}(r)}^{L_{{\rm max}}} \Phi(L)\,{\rm d}L,
\end{equation} 
where $\Phi(L)$ is the luminosity function of the objects under study and
$L_{\rm {min}}(r)=4\pi r^{2} S_{\rm {lim}}$, with $S_{\rm {lim}}$
being the flux limit of the sample under study.
In such a case the simple gravitational weights should be replaced by 
$w_{{\rm i}} \propto \phi^{-1}(r_{\rm i})\;r_{\rm i}^{-2}$, 
where the inverse selection function corrects (statistically) for the 
unseen portion of the luminosity function (for more details see 
K96 and references therein). 

A word of caution is due here. Since we sample
clusters up to 0.16 in redshift, we should have taken into account
possible evolutionary effects of the X--ray luminosity function. Instead, we
considered that the X--ray luminosity function (XLF hereafter) holds equally 
well for local and for distant objects which is clearly a simplification. 
Although this is still an open issue, there is recent evidence 
(Ebeling et al. 1996a; Nichol et al. 1997; Collins et al. 1997;
Ebeling et al. 1997a; Romer et al. 1997 and references therein) that
the cluster XLF does not evolve at least up to $z \sim
0.2$-0.3. Therefore we feel that using an non-evolving XLF is a safe
approximation at least for the distance range under study.
Had we used an evolving XLF, this would obviously have had a
direct impact on the accuracy of estimates of the contribution of
X--ray clusters to the soft XRB (depending on the degree and the
nature of the evolution).

The combination of the observed and
predicted peculiar velocity yields an estimate of the $\beta$-parameter via
equation (\ref{eq:lpt})  which is equivalent to:
\begin{equation}\label{eq:lpt1} 
{\mbox{\boldmath$u(r)$}}  = \frac{\beta}{4\pi\bar{n}}
{\mbox{\boldmath$D$}} = \beta {\bf V_{\rm {c}}},
\end{equation}
where ${\mbox{\boldmath$D$}}$ is estimated via equation (\ref{eq:dip}) and the
subscript c corresponds to cluster velocity.
The importance of using an accurate estimator for the mean density
of clusters is evident; there are alternative ways of constructing 
such an estimator and we refer the reader to K96 for details 
(Section 2.1). We verified that using different 
density estimators, the outcome is within 10 per cent of each
other (see also Davis $\&$ Huchra 1982). In what follows we just
estimate densities using equation (\ref{eq:self}).
Note that this method only generates a 
reliable estimate of $\beta$ if (i) the two vectors
${\mbox{\boldmath$u$}}$ and ${\mbox{\boldmath$D$}}$ 
are well--aligned (say within $25^{\circ}$) and (ii) that the effective depth 
of the sample, $R_{\rm max}$, exceeds $R_{\rm conv}$.
However, we should bear in mind that the very local 
contributions to the dynamics of the observer may not be represented in 
the cluster distribution, since these can be due to nearby galaxies, groups of 
galaxies or even voids in the local matter distribution. Such local effects 
can also be viewed as the cause of the observer-to-observer scatter in the 
linear biasing relation between the cluster and mass distributions 
(TB95; K96)

\subsection{Shot Noise Effects}
The sparseness with which the flux--limited cluster samples trace 
their underlying parent population introduce shot-noise 
(discreteness effects) in their dipole estimates which increase as a 
function of distance. The method of angular position reshuffling, 
used in K96 to estimate a similar effect for the $\it {IRAS}$ galaxies, 
is not suitable in this case.
Had we used the positional reshuffling method, we would have been
measuring  the convolution of the shot-noise introduced from the 
dilute sampling of the parent cluster population by its flux--limited 
subsample and
of a sort of intrinsic parent cluster population shot-noise dipole 
(under the false assumption that clusters are Poisson samplers of the 
underlying {\em mass} field).
In reality, since we will be inter-comparing cluster dipoles 
we are interested only in the former type of shot-noise which we
estimate according to the formalism developed in Strauss et al. (1992):
\begin{equation}
|{\bf D}|_{\rm {sn}}^{2}\approx \sum_{\rm i} \frac{1}{\phi_{\rm i} 
r^{4}_{\rm i}} \left[\frac{1}{\phi_{\rm i}}-1 \right] \;,
\end{equation}
where the sum is over all clusters in the flux--limited sample.
Evidently, the shot-noise dipole increases with depth and thus the 
amplitude of the flux--limited cluster dipole will also have a similar 
behavior, even beyond the true underlying dipole convergence depth. 
Therefore, in order to identify this convergence depth, a parameter which is
very important for a reliable determination of $\beta$ via equation 
(\ref{eq:lpt1}), 
we must correct the raw flux--limited dipole for the effects of shot-noise. 
Although the amplitude of the shot-noise dipole is easy to estimate, 
such a correction is not straightforward, because the shot-noise dipole 
direction is random and could point anywhere.

In the limit of uniform sampling we have from the central limit theorem that 
the three Cartesian components of the shot-noise dipole are equal and thus
$|{\mbox{\boldmath$D$}}|^{1d}_{\rm sn}=
|{\mbox{\boldmath$D$}}|^{3d}_{\rm sn}/\sqrt{3}$. 
Using a coordinate system in which the 1D shot-noise dipole is parallel to 
the direction of the shot-noise free dipole we have that 
$|{\mbox{\boldmath$D$}}|_{\rm corr} \approx |{\mbox{\boldmath$D$}}|_{\rm raw} 
\pm  |{\mbox{\boldmath$D$}}|^{1d}_{\rm sn}$.
We will correct for the effects of shot-noise using 
$|{\mbox{\boldmath$D$}}|_{\rm corr} \approx |{\mbox{\boldmath$D$}}|_{\rm raw} 
- |{\mbox{\boldmath$D$}}|^{1d}_{\rm sn}$ to provide maximum
discreteness correction and thus a rough lower limit to the estimated 
cluster dipole while we will also investigate the case of no shot-noise 
correction which should provide an upper limit to the estimated dipole.
The classical shot-noise correction,
$|{\mbox{\boldmath$D$}}_{\rm cor}| =
\sqrt{|{\mbox{\boldmath$D$}}_{\rm raw}|^{2}
-|{\mbox{\boldmath$D$}}^{3d}_{\rm sn}|^{2}}$
always produces results intermediate between the above limits and
therefore we will not consider it further.

Finally, an estimate of the misalignment induced by the shot-noise 
dipole on the true underline dipole direction is given by: 
$\delta\theta_{\rm sn} \approx {\rm arctan} 
\left(\sqrt{\frac{2}{3}}\frac{|{\mbox{\boldmath $D$}}|^{3d}_{\rm sn}} 
{|{\mbox{\boldmath $D$}}|_{\rm raw}} \right)$
(see also Juskiewicz, Vittorio \& Wyse 1990 and Lahav,
Kaiser \& Hoffman 1990 for dipole misalignment correction). 

\section{Samples and Simulations}
We now turn to the observational properties of the X--ray cluster samples
available at the present time, and the methods we use to simulate them. 
For more details on the observational data and motivation for this
work, see Guzzo (1995), Guzzo et al. (1995) and  de Grandi (1996).

\subsection{Properties of the Observed X--ray Cluster Samples}
In order that our simulations be as closely related as possible to
the available X--ray cluster data, we need to use an appropriate X--ray 
luminosity function and a realistic flux--limit.
We use the Schechter-like form for the luminosity function as provided by 
de Grandi (1996) with parameters given in Table 1:
\begin{equation}\label{eq:lf}
\Phi(L) = A \;\exp(-\frac{L}{L_{\ast}})\;L^{-\alpha}.
\end{equation} 
Here, $L_{\ast}$ is the characteristic luminosity, $A$ is the overall
normalization of the number--density,  and $\alpha$ is the
usual power--law index.
This luminosity function has been found to describe adequately a pilot
sample comprising 111 X--ray clusters having redshifts between 0.02 and
0.2 above the limiting flux of $3\,\times\,10^{-12}\,{\rm
erg\,s^{-1}\,cm^{-2}}$. 
The energy band for this application is 0.5-2 keV, instead of
the broader 0.1-2.4 keV, which we will be using. However, we can easily
convert results (fluxes or luminosities) from the hard band to the
band merely by using the conversion factor of order of 1.615, as suggested 
by de Grandi (private communication; see also section 3 of de Grandi 1996 
for details). Note that the above parameters of the luminosity function 
were derived from the original parameters of de Grandi (1996), by using 
the parametrised form of the Hubble constant. We prefer this conversion 
and consequently report results, which are $h$-dependent (see sections 4 
and 5).

\begin{table}
\centering
\caption[] {Parameters for the de Grandi (1996) luminosity function}
\tabcolsep 2pt
\begin{tabular} {ccc}
Parameters & Values & Units\\
$A$ & $2.3153\times10^{-6}$ & $(h^{-1}{\rm Mpc})^{-3}
(10^{44}\,h^{-2}\,{\rm erg}\,{\rm s^{-1}})^{\alpha -1}$ \\

\noalign{\smallskip}

$\alpha$ & $1.32^{+0.21}_{-0.23}$ & --  \\

\noalign{\smallskip}

$L_{\ast}$ & $0.66^{+0.218}_{-0.145} \times 10^{44}$ & $h^{-2}\,{\rm erg}\,
{\rm s^{-1}}$
\end{tabular}
\end{table}

It is necessary to implement a minimum distance criterion, because nearby
clusters are too extended to be identified by the present analysis software 
(SASS) of the {\small ROSAT} survey. Furthermore, the SASS
detection algorithm implementation leads to an underestimation of
the fluxes for all kinds of X--ray sources. In fact, both the pilot
sample of the ESO KP and the whole ESO KP survey make also use of an
alternative flux estimator (for both pointlike and extended X--ray emission),
namely the steepness ratio technique (for a detailed delineation, see
de Grandi et al. 1997 and references therein). Such a method greatly
improves the flux incompleteness initially produced by the SASS,
especially when the signal-to-noise ratio for the RASS sources is
small (i.e. when photon statistics is low).  
Nevertheless, we adopt a value of $R_{{\rm min}}\simeq
45 h^{-1}\,{\rm Mpc}$ ($z_{{\rm min}} \approx 0.015$ as in RASS) as
the minimum fiducial distance at which the sample is
complete. Although the maximum luminosity used for the pilot sample is
$1.01\,\times 10^{45}\,{\rm erg}\,{\rm s^{-1}}$, the most luminous
X--ray cluster identified within RASS yields the enormous value of
$6.2\,\times 10^{45}\,{\rm erg}\,{\rm s^{-1}}$ (Schindler et
al. 1995). This object appears to be incompatible with the luminosity
function (equations 6 and 11).
In any case, this is a preliminary
estimate of this function, so that our results which are based on it,
should be taken as indicative rather than definitive. 
Since the integration of this luminosity function is highly insensitive to
$L_{\rm {max}}$ for values well above $L_{\ast}$, we impose an upper limit 
of $1 - 1.2 \,\times 10^{45}\,h^{-2}\,{\rm erg}\,{\rm s^{-1}}$. 
The lower luminosity limit can be easily found in 
conjunction with $R_{{\rm min}}$ and the RASS flux limit.
Using the ESO Key Project flux limit, i.e.  
$S_{{\rm lim}}=3\,\times 10^{-12}\,{\rm erg}\,{\rm s^{-1}}\,{\rm cm^{-2}}$,
produces samples of few clusters, spanning a large redshift range, which 
therefore provide unreliable dipole estimates, principally because of the 
very large shot-noise uncertainties.  Since we are interested in producing 
theoretical predictions that can be compared with future, more sensitive 
surveys we use a deeper flux-limit, comparable to that of the 
RASS SGP project, set by Romer 
et al. (1994) and those by Collins et al. (1995) and Guzzo et al. (1995), i.e.
$S_{{\rm lim}}=1.5\,\times 10^{-12}\,{\rm erg}\,{\rm s^{-1}}\,{\rm cm^{-2}}$, 
which corresponds to $L_{{\rm min}} \approx 4\,\times 10^{41}\,h^{-2}\,
{\rm erg\,s^{-1}}$.
We do, however, have to hope that a survey with this deeper flux
limit would still be adequately described by the same luminosity function. 
Since, the pilot catalogue as given
above, reflects a quite uniform picture of the whole sampling space,
it is our reasonable hope that such an alteration of the flux limit will not 
affect the parameters of the luminosity function (L. Guzzo,  private 
communication).

\subsection{The simulations}
We now briefly describe the simulations we use to generate mock samples
of galaxy clusters. For a more detailed description, see Borgani et al. 
(1995).

Our prescription is based on an optimized version of the Truncated
Zel'dovich approximation (Zel'dovich 1970; Shandarin \& Zel'dovich 1989;
Coles, Melott \& Shandarin 1993). In this approximation, known as TZA,
particles move along straight lines in response to the initial
gravitational potential generated by a smoothed version of the
initial density field. This describes the dynamics quite accurately up to the
shell-crossing point, which is suppressed by the initial smoothing.
In the simulations we use here,
density and velocity potentials are reconstructed on
$256^{3}$ grid points in a box of $L=960 h^{-1}\,{\rm Mpc}$ aside. 
It has been shown that TZA
accurately locates clusters when  $\sigma_{8} < 1$ (Borgani et al 1995; 
Moscardini et al. 1996). Indeed, this is the case for the model we
will be dealing with in this paper, a cold + hot dark matter
model (CHDM hereafter) with $\Omega_{\circ}=1$ for the total density, 
$\Omega_{{\rm hot}}=0.3$ for the hot component and $\Omega_{{\rm b}}=0.013 
h^{-2}$ for the baryonic fraction (Reeves 1994). 
We also use $h=0.5$ for the dimensionless
Hubble parameter and $\sigma=0.78$ for the rms fluctuation amplitude
within a top-hat sphere of radius of $8 h^{-1}\,{\rm Mpc}$. 
In order to reduce the amount of shell 
crossing we filter the short-wavelength modes of the linear power spectrum 
using a Gaussian filter with radius $R_{{\rm f}}=3.3 h^{-1}\,{\rm Mpc}$. 
The model is normalized to COBE as in  Gorski et al. (1994). 
We use this single model that has been shown to provide a
good description of the cluster correlation and large-scale properties 
(cf. Klypin \& Rhee 1994; Borgani et al. 1997).

In TB95 it was shown that using large simulation volumes the variance
in the dipole estimates is dominated by the observer-to-observer scatter 
within the same realization and not by the different realizations of
the same dark matter model. Therefore, in what follows, we will be
using only a single realization of the model.

\subsection{Cluster Sample Extraction}
The simulation described above allow us to identify clusters 
with peaks of the dark matter density field and a subset of these
peaks are used to generate a simulated cluster sample. In our previous
analysis of cluster samples (K96), we assumed that there was a 
correlation between the height of the density peak and the optically 
defined richness of a cluster. In this work, we assume a
similar correlation between the height of the peak and the cluster X--ray
luminosity. To simulate realistic flux--limited samples of X--ray clusters 
we need the number density of their parent population
which can be converted into a threshold for peak selection.
This is used to generate the ``parent'' volume--limited catalogue of
clusters. Then by imposing a flux limit, and the shape of the {\small
ROSAT} selection function we extract realistic X--ray cluster 
samples.

\subsubsection{``Parent'' Cluster Catalogue (C)}
By integrating the luminosity function from the proper lower to
the upper luminosity limit, we can get an estimate of the mean number density
of the ``parent'' distribution of X--ray clusters:
\begin{equation}\label{eq:den}
\bar{n}_{\rm c}=\int_{L_{{\rm min}}}^{L_{{\rm max}}} \Phi(L) \; {\rm d}L.
\end{equation}
Taking the lower and the upper bound given in section 3.1, we obtain 
$\bar{n}_{\rm c} \approx 3.3\,\times10^{-5}\,h^{3}\,{\rm Mpc^{-3}}$.
Thus the number of ``parent'' clusters
included within the cubic box is $N_{\rm c} \approx 28\,900$ and the
number within an  inscribed sphere of radius $480 h^{-1}\,{\rm Mpc}$ 
is $\approx15\,100$ which corresponds to a mean intercluster separation of 
$\bar{d}_{\rm c} \approx 31.3h^{-1}\,{\rm {Mpc}}$, which is characteristic
of the APM type of clusters (Dalton et al. 1994).

\subsubsection{The mock `ROSAT' sample ($\equiv R$)}
To simulate a {\small ROSAT}-like flux--limited X--ray cluster sample 
we use the selection function, defined in equation (\ref{eq:self}), from 
which we can predict the corresponding number of X--ray clusters  
above the flux limit, $S_{{\rm lim}}=1.5\,\times 10^{-12}\,{\rm
erg}\,{\rm s^{-1}}\,{\rm cm^{-2}}$, lying within a shell between $r$
and $r+\Delta r$, via:
\begin{math}
N(r)=4\pi r^{2}\bar{n}\phi(r)\Delta r.
\end{math} 
Fig. 1 shows the resulting $\phi(r)$ as a
function of distance. Notice the very long tail extending out to large
distances.
The maximum of $N(r)$ turns out to be around $z\approx 0.11$, 
consistent with that predicted by the RASS (L.Guzzo, private communication).
These considerations lead to a value of $N_{{\rm c}}\approx
1300$ clusters within our limiting redshift of $z\approx 0.16$. We are 
therefore in perfect aggreement with the original survey target list of about 
700-800 southern clusters (within $z\approx 0.2$). 

Finally we perform a consistency check, that of recovering
from our simulated flux--limited samples the observed $N-S$ 
distribution (de Grandi 1996) which is well-described by a power law of 
the form
\begin{equation}\label{eq:chi}
N(>S_{{\rm i}})\approx A\,S_{{\rm i}}^{-\alpha},
\end{equation}
with $\alpha=1.21\pm 0.15$ and normalization constant 
$A=11.34^{+1.80}_{-2.14}\,{\rm {sr}}^{-1}\,(10^{-11}\,{\rm
{erg\,s}}^{-1}\,{\rm {cm}}^{-2})^{\alpha}$. 
Using 500 random observers, a flux-limit comparable to
that of de Grandi (i.e. 
$S_{{\rm lim}}=3\,\times 10^{-12}\,{\rm erg}\,{\rm s^{-1}}\,{\rm cm^{-2}}$)
and fitting equation (\ref{eq:chi}), we find $\alpha=1.22\pm 0.05$, $A=11.7\pm
1$ in units as before and in excellent aggrement with 
de Grandi (1996). 

\subsubsection{A `test' sample (T)}
This test sample is constructed in order to study the effects of missing 
nearby clusters which are too extended to be picked up by the 
SASS identification procedure (inability of the standard analysis software 
system to define them as such, simply because it is point source
detection algorithm, therefore it can not deal properly with extended
X--ray emission; but see also section 3.1 for additional flux
correction). We use exactly the same requirements as for the R-sample,
but now we adopt $R_{{\rm min}}=7.5 h^{-1}\,{\rm Mpc}$ instead.  
The number of objects per simulation is only slightly
affected, since we include the few clusters that lie between
7.5$h^{-1}\,{\rm Mpc}$ and 45$h^{-1}\,{\rm Mpc}$ (according to their mean 
separation we expect $\approx$ 10 clusters within these limits). 
This particular lower limit is dictated by the procedure with which we 
have identified the observers; all local properties are defined by using a 
top-hat sphere of radius 7.5$h^{-1}\,{\rm Mpc}$ (for more 
details see TB95).
Although such an inclusion increases insignificantly the total number of 
clusters, it does influence all dipole properties to a great extent, 
as expected from the $1/r^2$ law of gravity and as we will quantify
in our subsequent analysis.

\subsection{Observer selection}

From a starting list of 5000 random observers, identified on the grid, we 
will be selecting appropriate subsets with different characteristics
depending on which particular issue we want to study.

Firstly we select only those observers for which their cluster dipole 
amplitude, on large scales ($\sim 180h^{-1}\,{\rm Mpc}$), exceeds their 
shot-noise dipole. This results in 1500 such observers, which we call random 
observers. 
We imposed this last criterion since we are interested in studying dynamically
`active' regions, for which the dipole is well defined and a relative good 
diagnostic of the acceleration field.

As a second subset we are interested in defining LG-like observers.
Imposing the usual LG requirements (similar peculiar velocity
and local density contrast; cf. Borgani et al. 1995; Moscardini et al. 
1996) we generate an insufficient number of observers to allow a
meaningful statistical analysis to be performed.
A more useful constraint for studying the large-scale characteristics of 
the cluster distribution (which we also used in K96) is furnished by the 
the structure of the observed Abell/ACO cluster dipole, which is of a 
two-step form with a significant contribution from large scales (cf. 
Scaramella et al. 1991; Plionis \& Valdarnini 1991; 
Branchini \& Plionis 1996).
In other words, we will just assume that our ``parent'' X--ray cluster 
distribution has similar dipole amplitude and shape as the Abell/ACO 
optical clusters. 
Branchini \& Plionis (1996) starting form the redshift space cluster 
distribution they recovered cluster true distances and peculiar
velocities via a dynamical algorithm (cf. Yahil et al 1991; 
Strauss et al 1992).
Their mean reconstructed real space cluster dipole has the
following features.
\begin{enumerate}
\item $V_{\rm C} \approx 2400\,\pm\,800 \;{\rm km}\,{\rm s^{-1}}$
(2$\sigma$ range) and
\item the usual two-step shape, with a final convergence at 
$\sim 180h^{-1}\,{\rm Mpc}$.
\end{enumerate}
This  requirement reduces the number of available observers in 
the simulation to 200, observers which we will hereafter call 
LG-like observers.

\subsection{Weighting Schemes}
We finish this section by discussing the appropriate weighting schemes, to be
used in equations (\ref{eq:mon}) and (\ref{eq:dip}), for the various samples.
The C sample, being the ``parent'' population, is volume limited and thus does 
not suffer from selection effects, so we adopt the direct weight:
$w_{\rm i}=r_{\rm i}^{-2}$. 
For the R,T-samples, which are flux--limited, we use
$w_{{\rm i}}\approx \phi^{-1}(r_{{\rm i}})\,r_{{\rm i}}^{-2}$ to correct 
for the missing objects. Note that this weighting scheme assumes that the 
unseen objects follow the same luminosity function and that they are 
spatially, strongly correlated with the observed objects.
Note that for the T-sample we use $\phi(r)=1$ for distances between 
$7.5h^{-1}\,{\rm Mpc}$ and $45h^{-1}\,{\rm Mpc}$.

\section{Analysis and Results}
We compute the moments of the cluster distribution for all 3 samples 
(C, R, T), in bins of $20h^{-1}\,{\rm Mpc}$ width.
We sum the contributions from all objects up to
$R_{{\rm max}}=480h^{-1}\,{\rm Mpc}$ using periodic boundary conditions. 

\subsection{Monopoles}
We plot in fig. 2 the average random observer monopoles,
for all three sets of clusters calculated according to 
equation (\ref{eq:mon}), which as expected increase linearly with $r$. 
For the flux--limited (R and T) samples, however,
we observe a systematic early flattening at $\sim 100h^{-1}\,{\rm
Mpc}$, which induces
an overall underestimation of the ``parent'' population (C) monopole, although 
they are boosted up by using the appropriate correction factor 
$\phi^{-1}(r_{{\rm i}})$.
Their relative underestimation of the ``parent''-population monopole, measured 
at depths $R\geq R_{{\rm conv}}$, is $\delta M \approx$ 10 and 20 per cent 
respectively for the T and R samples. This implies that the mean density of 
X--ray clusters, calculated using the flux--limited sample monopole (via 
$\bar{n}=M/4 \pi R$), would be an underestimate of their true underlying 
mean density, which would then provide an artificially lower estimate of 
$\beta$ in equation (\ref{eq:lpt1}). 
We note that, up to $100h^{-1}\,{\rm Mpc}$, we expect on average $\approx
50-60$ clusters.

\subsection{Dipole as a function of distance}

To study the similarities of the different sample dipoles shape we
use the usual correlation coefficient analysis (see equation 17 of  K96).
Furthermore to examine the accuracy with which the ``parent'' cluster 
distribution is traced by the flux--limited R and T samples, 
we measure their corresponding velocity (dipole) fluctuations as follows.
\begin{equation}\label{eq:velf}
\Delta V_{\rm{i,C}}(R) = \frac{V_{\rm{i}}(R)-V_{\rm{C}}(R)}{V_{\rm{C}}(R)}
\end{equation} 
The subscript i corresponds to either R or T-samples and the subscript C
corresponds always to our ``parent'' cluster sample 
(see also equation 18 of K96).
Evidently, if the ``parent'' cluster population is adequately
traced by flux--limited samples, we would expect to find $\Delta V_{{\rm
i,C}}(R)\approx 0\; \forall\; R$, while in the case of missing 
contributions we would obtain  $\Delta V_{{\rm i,C}}(R) < 0$.

As we anticipated for the case of flux limited samples (section 2.3),
we have found that in many cases the raw dipoles (before correcting
for the shot-noise effects) diverge as a function of depth beyond the
depths were their parent 3D cluster dipole has already converged. 
Therefore, we will present results based on both shot-noise correcting
techniques discussed in section 2.3, although we expect that for some
observers such a correction will underestimate their true dipole
amplitude.

\subsubsection{Random observers}
In most of the cases the ``parent'' sample (C) dipole amplitude is constantly
higher than that of the T-sample. Note that, in both these cluster sets, 
the `local region' up to $45h^{-1}\,{\rm Mpc}$ is included. The dipole
shapes are quite similar up to $R_{{\rm max}}$, which is evident from
their relative shape correlation coefficient being $\approx
0.84\,\pm\,0.13$. 
However, the dipole amplitude of the C sample is underestimated by the 
flux--limited sample, i.e. $\Delta V_{{\rm T,C}}(R)\approx -0.18$ and -0.1
for the shot-noise corrected and uncorrected case resepectively.

Comparing the C and R dipoles we find 
$\Delta V_{{\rm R,C}}(R)\;\mincir -0.3$ and -0.2
for the shot-noise corrected and uncorrected case resepectively, while
their relative dipole-shape correlation coefficient is $0.81\,\pm\,0.12$.
This implies that in many occasions the C dipole is
dominated by substantial `local' contributions, within 
$(50-60)h^{-1}$ Mpc which the R dipole inevitably miss 
by construction and thus underestimates even further the ``parent'' 
population dipole. 

Examining the dipole amplitudes of the three cluster samples, 
we identify the following categories in decreasing order of frequency.
\begin{enumerate}
\item $V_{{\rm C}}>>V_{{\rm T}}\geq V_{{\rm R}}$ in 60 per cent of the
observers.
\item $V_{{\rm C}}\approx V_{{\rm T}}\approx V_{{\rm R}}$ (within 10 per
cent) in 30 per cent of the observers and
\item $V_{{\rm R}}>V_{{\rm C}}$ in 10 per cent of the cases.
\end{enumerate}

In fig. 3, we plot dipoles for the three samples as a function of radial
distance. Panels (a) and (b) are representative of the first category, 
panel (d) of the second category and panel (c) is of the third case. 
Most of the R-dipoles
underestimates the ``parent'' population dipole up to a distance of 
$\approx 100h^{-1}\,{\rm Mpc}$,
where they actually start gaining amplitude. The final convergence comes
at $R_{{\rm conv}}\approx 200h^{-1}\,{\rm Mpc}$.  

We also plot in fig. 4, the relative velocity fluctuations 
($\Delta V$). Note that, in the C versus T case the velocity 
fluctuation, $\Delta V_{{\rm C,T}}(R)$, convergence to its final value
at $\approx 120h^{-1}\,{\rm Mpc}$, whereas the situation is entirely
different for $\Delta V_{{\rm C,R}}(R)$, as expected from the fact that
the local region ($< 45h^{-1}\,{\rm Mpc}$) has been excluded in the R
sample. Here
at depths below $100h^{-1}\,{\rm Mpc}$, we observe a very low value
$\Delta V_{{\rm C,R}}(R) \approx -0.6$, which reaches its final value 
($\approx -0.3$) at $R\approx 180h^{-1}\,{\rm Mpc}$. 

It is important to note that, had we not corrected at all for the
shot-noise effects (section 2.3) we would have found lower
underestimates of the parent cluster dipole amplitude (i.e. $\Delta
V(R) \approx -0.1 \sim -0.2$) as expected from the fact that our
shot-noise correction is maximal (see discussion in section 2.3).  
In general, this indicates that the underestimation of the true
underlying cluster dipole amplitude by the flux--limited samples is
counteracted at some level by the artificial dipole amplitude
enhancement due to the shot-noise effects. The shot-noise correction
is however very important in order to identify the true dipole
convergence depth. 

\subsubsection{LG--like observers}
Using equation (\ref{eq:velf}) we find similar $\Delta V$ values as in the 
random observer case (i.e. $\Delta V_{{\rm T,C}}(R_{{\rm
conv}})\approx -0.16$ and  $\Delta V_{{\rm R,C}}(R_{{\rm
conv}})\approx -0.3$), which again represent an underestimate
of the C-dipole. However, the correlation coefficients of the dipole shapes 
are systematically larger in this case than for the random observers. 
We find $0.9\,\pm\,0.1$ for the C versus T samples and
$0.87\,\pm\,0.09$ for the C versus R ones. It is clear that, the
three dipoles are in good agreement, as far as their shapes are concerned. 

We can break down the  behavior of the dipole amplitude for the
LG-like observers into the same categories as for the random observers
case. Only slight changes occur in the relative frequencies with most
notable the slight increase of the observers for which their
flux--limited samples provide good estimates of the ``parent''-population
cluster dipole (C) ($\sim 40$ per cent of total). We do not present
plots for the LG-like observers due to their qualitative similarity
with those of the random observers (figs 3 and 4). 

Finally, we report that for both sets of observers we have also
applied the classical shot-noise correction of section 2.3.
Although we do not show any plots for reasons of
brevity, results clearly suggest that whatever the adopted method for
such corrections, both flux--limited cluster samples (R, T)
underestimate the ``parent'' (3D) dipole by at least 10 per cent
independently of the observer.

\subsection{Dipole alignment}
In fig. 5 we present the relative dipole misalignment angles, $\Delta \theta$,
between the three sample dipoles, averaged over the ensemble of all observers. 
We define $\Delta \theta_{{\rm C,T}}$ to be the misalignment angle
between the dipole vectors of the C and T samples, and 
$\Delta \theta_{{\rm C,R}}$ between C and R.
Panel (a) shows the $\Delta \theta$ as a 
function of distance for the random observers while panel (b) 
for the LG-like observers.

For the random observers, we obtain mean values of 
$\Delta \theta_{{\rm C,T}}\sim 18^{\circ} \pm\,15^{\circ}$ and
$\Delta \theta_{{\rm C,R}} \sim 47^{\circ} \pm\,30^{\circ}$. 
The above numbers have been measured at $R_{{\rm conv}}$. 
The big difference of $\approx 30^{\circ}$ in the mean misalignment angles
between $\Delta \theta_{{\rm C,T}}$ and $\Delta \theta_{{\rm C,R}}$ can
be attributed to the fact that the R dipole ignores (by construction) 
the local region ($<45h^{-1}\,{\rm Mpc}$). 
This again shows how important the `local region' actually is in determining 
the cluster dipole.
From panel (a) of fig. 5, it is clear that the misalignment angle
between the C and T sample dipoles converge to its final value
at $\sim 100h^{-1}\,{\rm Mpc}$ while between
the C and R-dipoles this scale is $\sim 200h^{-1}\,{\rm Mpc}$. 
For distances well below this limit, we get $\Delta\theta \sim (60^{\circ} - 
70^{\circ}$).

As seen from panel (b), the situation is greatly improved, when we use
the LG-like observers, since by construction their dipole properties are not
dominated only by the `local region'. The large misalignment angles are 
reduced rapidly and converge to a significantly lower value at $\sim 150 
h^{-1}\,{\rm Mpc}$. We obtain $\Delta \theta_{{\rm
C,T}}\approx 17^{\circ}\pm 11^{\circ}$ with a median of $18^{\circ}$ and
$\Delta \theta_{{\rm C,R}}\approx 30^{\circ}\pm\,20^{\circ}$ with a
corresponding median of $31^{\circ}$. 

Note, finally, that we did not apply any correction for shot-noise 
effects in the dipole direction analysis. A shot-noise correction is
expected to reduce the dipole misalignment angle by at least
$10^{\circ}$.

\subsection{ A Cluster Bias Estimate}
Using equation (\ref{eq:lpt1}) and the input value of 
$\Omega_{\circ}\,(=1)$, given by the model we simulate, we can estimate 
the cluster bias parameter, via: 
\begin{equation}
\label{eq:bias}
b_{\rm c} = \frac{1}{N}\sum_{\rm
i=1}^{N}\frac{|{\mbox{\boldmath$V$}}|_{\rm c,i}
(R)}{|{\mbox{\boldmath$u$}}|_{\rm i}(R)},
\end{equation}
where $N$ is the total number of observers used and $R \ge R_{\rm conv}$.
The small subscript c indicates cluster velocity for the three
samples (see section 3.3 and equation 7), whereas i shows the
specific observer used in the summation.
A more elaborate version of this procedure was tested and found to provide
a reliable estimate of $b_{\rm c}$ in K96. Using equation (\ref{eq:bias}) 
and those LG-observers for which their peculiar velocity vector lie within 
$\sim 25^{\circ}$ of their measured cluster dipole we obtain for the 
``parent'' sample: 
\begin{equation}
b_{\rm c}=4.8\pm1.5\;,
\end{equation} 
a value which is very similar to that derived by the ratio of the 
cluster to dark-matter variances (measured on scales $>$ $50h^{-1}$ Mpc). 
However the method of equation (\ref{eq:bias}) produces a large 
observer-to-observer
scatter (6 to 8 times larger than in variance method) which is the outcome of 
the fact that the very local contributions to the dynamics of some observers 
may not be represented in the cluster distribution, since these 
can be due to the very local inhomogeneities in the matter distribution. 
 
The underestimation of the ``parent''-population dipole amplitude by the 
flux--limited samples will inevitably reflect itself on an artificially
lower bias parameter, given by equation (\ref{eq:bias}). 
In fact we obtain $b_{\rm c} \approx 3.95\pm 1.6$ and $3.45\pm 1.4$ for the T 
and R samples respectively.

\subsection{Summary of Moment Analysis}
The flux--limited cluster samples underestimate the X--ray ``parent''
population dipole. All dipoles are in good agreement (within 10 per
cent), in about only one in three cases.  
The main effect of imposing a flux-limit is a general loss of dipole
amplitude, which amounts to 10/20 per cent for the T sample and up to
20/30 per cent for the R sample (depending on whether one corrects or
not for shot-noise effects). Note that the above results are 
independent of the two methods used (discussed in section 2.3) to correct 
for the shot-noise effects. 
The relative difference between the two flux--limited 
samples (which can reach a maximum amplitude of  $\sim 15$ per cent) 
is due to the exclusion of the `local region' from the R sample, which  
corresponds to the inability of the RASS to sample the local extended
X--ray cluster sources. The misalignments between the flux--limited
sample dipoles and that of the ``parent''-population are quite large,
ranging from 20$^{\circ}$ to $47^{\circ}$ on average (for the T and R
samples respectively). 

If we restrict our analysis to those observers that have a ``parent'' cluster 
population that exibits a similar dipole to the observed Abell/ACO
cluster case (i.e. having significant contribution from large scales; 
cf. Branchini \& Plionis 1996), the two flux--limited samples have dipole 
directions in much better 
aggreement with that of the ``parent'' cluster dipole ($\Delta\theta \sim 
17^{\circ}$ and $30^{\circ}$ respectively for the T and R samples). The dipole
amplitude underestimation by the flux--limited samples is however of a similar 
amplitude as in the random observers case.

Note, importantly, that the underestimation of the ``parent'' cluster dipole 
amplitude by these flux--limited samples compared would lead, using 
equation (\ref{eq:lpt1}), to an incorrectly higher estimate of the 
$\beta_{\rm c}$ parameter and thus to either an artificially higher 
value of $\Omega_{\circ}$ for a specific value of $b_{\rm c}$ or to an
artificially lower value of $b_{\rm c}$ for a specific value
of $\Omega_{\circ}$.
\section{X--ray cluster contribution to XRB}
\subsection{A brief overview}
Since it was brought forward by Giacconi et al. (1962), the XRB 
has been something of an enigma for theoretical cosmology.
Its origin and its major constituents are still debatable nowadays, and it
is still unclear how much of it is contributed by resolved sources
and how much by a discrete component (see review by Zamorani 1993 as well as
Hasinger et al. 1993; Hasinger 1992; Hasinger 1995).
The diffuse hypothesis (thermal breemstralung hot plasma with $T\approx
40\,{\rm keV}$) has largely been abandoned, 
because such a hot component would generate
severe distortion on the CMB spectrum, which is not detected
by COBE. The alternative solution of assuming that the XRB is made up of
the integrated X--ray flux from discrete objects is more attractive, but as
yet no specific class of sources has been found with a
spectrum identical to that of the XRB.

Setting the spectral difficulties to one side, we shall attempt to
quantify the contribution to the XRB from clusters of galaxies.
Many similar attempts have been made for various other sources
(active galaxies, QSOs) as well as clusters, in both soft
($E\,\leq\,2$ keV) and hard ($E>2$ keV) energy bands; see
Shanks et al. (1991) (and references therein), Blanchard et al. (1992), 
Georgantopoulos et
al. (1993), Boyle et al. (1994), Roche et al. (1995) (and references
therein) for the QSO contribution to the soft XRB, 
Lahav et al. (1993) (and references
therein) discuss the contribution from galaxies, Anvi (1978) and Maccacaro
et al. (1984) have discussed the case for AGN and Piccinotti et
al. (1982), Gioia et al. (1990), McKee et al. (1980), Schwartz (1978),
Henry et al. (1992), Soltan et al. (1995), Kitayama \& Suto (1996),
Kitayama, Sasaki \& Suto (1997) and Oukbir, Bartlett \& Blanchard
(1997) are mostly concerned about X--ray clusters.

We will attempt to estimate the contribution of X--ray clusters,
having a X--ray luminosity function given by equation (\ref{eq:lf}) to
the soft XRB using two simple methods (see Maccacarro et al. 1984, their 
section 4), not worrying for the different cosmological backgrounds
that could affect the detailed modelling of such contribution
(cf. Kitayama \& Suto 1996; Kitayama et al. 1997). Note that the
present application holds only for a flat model with vanishing
$\Lambda$ term, since this is the background that de Grandi
(1996) has assumed in deriving the relative parameters of the ESO-KP
X--ray luminosity function. 

\subsection{Luminosity Function Method}
First we compute the contribution of X--ray clusters either in terms of
flux per steradian or in terms of luminosity density
(${\rm erg}\,{\rm s^{-1}}\,h\,{\rm Mpc^{-3}}$). In other words, we 
compare either intensities of clusters and XRB, or volume emissivities at 
a given energy band. The contribution of clusters is estimated integrating
the product $L\,\Phi(L)$:
\begin{equation}\label{eq:Jc}
J_{{\rm c}}=\int_{L_{{\rm min}}}^{L_{{\rm max}}} 
L\,\Phi (L)\,{\rm d}L\,,
\end{equation}
where $L_{\rm min}\approx 4 \times 10^{41}\,h^{-2}\,{\rm erg}\,{\rm
s^{-1}}$ and $L_{\rm max}\approx 10^{45} \,h^{-2}\,{\rm erg}\,{\rm
s^{-1}}$ (see section 2.2).
The quantity $J_{{\rm c}}$ expresses the integrated cluster volume 
emissivity.
Since there is no clear picture which describes the corresponding volume
emissivity of the XRB in such a soft band, we will simply be using the
well-defined intensity of the XRB in the 1-2 keV band, namely     
$I_{{\rm X}}\approx 1.25\,\times 10^{-8}\,{\rm erg}\,{\rm s^{-1}}\,{\rm
cm^{-2}}\,{\rm sr^{-1}}$ of
Hasinger et al. (1993), as the basis of our estimate. 
Assuming values for the spectral index of the soft XRB in the usual
range of 1-1.2, we first convert the latter value
in the monochromatic band (at 1.5 keV) and then we use a conversion
factor to transform it to the 0.1-2.4 keV one.
Such a calculation yields $I_{{\rm X}} \approx
4\,\pm\,0.2\,\times\,10^{-8}\,{\rm erg}\,{\rm s^{-1}}\,{\rm cm^{-2}}\,{\rm sr^{-1}}$
(0.1-2.4 keV), for the XRB intensity in the usual ROSAT band. 
Evidently, we should compare intensities with luminosities in identical
bands and for this purpose we either convert the cluster volume emissivity
in flux, or the intensity of the XRB in light density units. The former
requires an accurate knowledge of the spectral index of X--ray clusters,
for the conversion to be valid. 
We will assume a typical value for the spectral index of X--ray clusters in the
soft band of the order of 0.4 (photon index 1.4).
The relation between volume emissivities and intensities
(Schwartz \& Gorsky 1974; Anvi 1978; Soltan et al. 1995) is given by 
\begin{equation}\label{eq:Io}
I_{{\rm i}}=\frac{c\,J_{{\rm i}}}{4\,\pi\,H_{\circ}} \int_{0}^{z_{{\rm
max}}}\;\frac{(1+z)^{1-\alpha_{{\rm i}}}}{(1+z)^{3}\,
\sqrt{1+2q_{\circ}z}}\,{\rm d}z.
\end{equation}
The subscript i corresponds to the objects under study; one can use 
equation (16) interchangeably for 
$I_{{\rm i}}$ or $J_{{\rm i}}$. The look-back integral depends on  
parameters such as $H_{\circ}, q_{\circ}$ and $\alpha_{{\rm i}}$ (a zero 
cosmological constant is assumed). Varying $q_{\circ}$ in the range between 
0.1 and 0.5 does not affect the the look-back time integral by more
than 10 per cent. The cluster contribution to the soft XRB, if
$q_{\circ}$ is in the above range, is therefore well within the
uncertainties quoted throughout this section. Furthermore the 
dependence on $H_{\circ}$ drops out from the final result. 
Using the spectral index of the XRB given above, we can solve for 
$J_{{\rm X}}$. The
upper limit of the integral presents the maximum distance up to which,
sources generate the XRB. Values between 3, 4 and $\infty$ do not
influence the final results by more than 3 per cent. In what follows we
will be using $z_{{\rm max}}=4$ and $q_{\circ}=0.5$. 
We then obtain from equation (16) that $J_{{\rm X}}\approx 4.1\, 
\pm\,0.1\,\times\,10^{39}\,{\rm erg}\,{\rm s^{-1}}\,h\,{\rm Mpc^{-3}}$
(0.1-2.4 keV).
Adopting the corresponding value for the clusters (equation 15), we
can express the fractional contribution of X--ray clusters to the total
of the soft XRB as 
\begin{equation}\label{eq:C}
{\cal C}_{0.1-2.4\,{\rm keV}} = \frac{J_{{\rm c}}}{J_{{\rm X}}}\approx
(9\pm 1)\,\%,
\end{equation}
which is independent of the value of $H_{\circ}$.

We can obtain another estimate of ${\cal C}_{0.1-2.4\,{\rm keV}}$, starting 
with the formalism of Schwartz \& Gorsky
(1974), using equation (16) and the spectral intensity of the XRB for
energies between 1 keV and 21 keV and then solving for $J_{{\rm X}}$.
The spectral intensity is expressed as:
\begin{equation}\label{eq:Ie}
I(E)\approx 8.5\,E^{-0.4}\,\frac{{\rm keV}}{{\rm keV}\,
{\rm cm^{2}}\,{\rm s}\,{\rm sr}},
\end{equation}
for $1\leq E\leq 21$ in keV's. It is apparent though, that such a scheme
does not cover the lower limit of our soft band (at 0.1 keV). 
We can circumvent, however, this difficulty by using the following
alternatives. 
The first is to extrapolate the shape of the spectrum at low energies, so as 
to match our lower bound, and the second is to perform
calculations in the perfectly allowed (1-2 keV) band, which we later
convert in the soft ROSAT one as before. We choose the second method, since it
does not violate the energy range and it appears to be more realistic.
However, the former method produces figures always within
$\pm\,1\sigma$ of the latter.
We caution the reader that there is no well-known $I(E)$ for the band 
we are interested in and this energy range is mostly unexplored in terms of 
spectrum. Taking again the same set of parameters as
before ($H_{\circ}, q_{\circ}, \alpha_{{\rm X}}, I_{{\rm X}}$), we obtain 
$J_{{\rm X}}\approx 3.95\,\pm\,0.1\,\times
10^{39}\,{\rm erg}\,{\rm s^{-1}}\,h\,{\rm Mpc^{-3}}$ (0.1-2.4 keV), which 
is in good agreement with the value previously
computed and which gives
${\cal C}_{0.1-2.4\,{\rm keV}} \approx (10\pm 1)\,\% \,.$
Note that the errors of both estimates reflect
the uncertainty of the choice of the spectral index of the soft XRB, 
being used in equation (16) (cf. Soltan et al. 1995). 

A further estimate of ${\cal C}_{0.1-2.4\,{\rm keV}}$
is obtained using an estimate of the spectral index of clusters 
of the order of $\alpha_{{\rm c}}=(0.4-0.5)$, as in Maccacaro et
al. (1988), and using the value of $J_{{\rm c}}$ from equation
(\ref{eq:Jc}) in the 0.1-2.4 keV band, we can then solve equation
(\ref{eq:Io}), this time for the  total intensity for X--ray clusters,
$I_{{\rm c}}$, in the same energy range.  We find that $I_{{\rm c}}
\approx 4.5\,\pm\,0.1\,\times 10^{-9}\,{\rm erg}\,{\rm s^{-1}}\,{\rm
cm^{-2}}\,{\rm sr^{-1}}$ (0.1-2.4 keV).  
By comparing this with the previous estimate of $I_{{\rm X}}$
(0.1-2.4 keV), we find 
\begin{equation}\label{eq:C2}
{\cal C}_{0.1-2.4\,{\rm keV}} =\frac{I_{\rm c}}{I_{\rm X}} \approx 
(11 \pm 1) \,\%
\end{equation}
consistent with the previous estimates.
For all the above estimates, we have assumed that the
spectral indices of both clusters and XRB are not subject to any 
change when passing from one energy band to another and although a few
such assumptions have been used, this framework does yield consistent
results. 

\subsection{Number-Flux Distribution Method} 
A different approach is used here, which is independent of the value of 
$q_{\circ}$, that enters in the evaluation of the look-back integral.
We assign a random luminosity to each simulated cluster using the 
X--ray luminosity function of equation (\ref{eq:lf}), which we then
convert to a flux using its distance. For this, we use the constant multiplier 
of de Grandi (1996) to transform  0.5-2 keV fluxes (or luminosities)
to the broad 0.1-2.4 keV band (as discussed in section 2.2). Note that
the particular CHDM simulation we use is irrelevant to this analysis;
we could have even used a Monte-Carlo cluster distribution provided
that the $N(r)$ distribution, predicted by the X--ray cluster selection
function, would be preserved.
We plot in fig. 6 the distribution of X--ray luminosities as a
function of distance. The solid line corresponds to the imposed
flux limit of $1.5\,\times 10^{-12}\,{\rm erg}\,{\rm s^{-1}}\,{\rm
cm^{-2}}$, for any given distance.
The plot shows results for a single observer; we have verified
that this behavior of the distribution is entirely independent of the
observer position.

Having obtained the fluxes of the $\sim 1300$ X--ray-like clusters, we 
simply add their individual contributions towards the total cluster X--ray 
emission in the 
corrected  soft band. To estimate the scatter due to sampling effects
we use 500 random observers from 
the initial ensemble and we repeat the procedure for each number-flux
relation. We report results, as usual, as the mean and the $\pm\,
1 \sigma$ errors. We find that, $I_{{\rm c}}\approx 4.25\pm 0.3 \times
10^{-9}\,{\rm erg}\,{\rm s^{-1}}\,{\rm cm^{-2}}\,{\rm sr^{-1}}$ (0.1-2.4 keV).
We  compare intensities of clusters and XRB as computed in section
5.2. We obtain ${\cal{C}}_{0.1-2.4\,{\rm keV}} \approx
10.5\pm 1$ per cent for the contribution of the discrete set of these
clusters. Alternatively, we can again assume equation (\ref{eq:Io}) 
and the same spectral index for clusters, in order to convert fluxes per 
unit area to volume emissivities. This approach yield identical results.

However, since we have used our simulated catalogue which is limited to a 
radius of $480h^{-1}\,{\rm Mpc}$, we need to take into account the contribution
of the extra $\sim$ 300-400 clusters predicted by the 
X--ray cluster selection function to lie $z\;\magcir 0.16$, which we
estimate to be no more than 1.5 per cent of the previously quoted value. 
Therefore this estimate is in very good agreement with our previous
findings (see equations \ref{eq:C} and \ref{eq:C2}). 

We, therefore, quote the contribution of clusters to the soft XRB, as 
an average of the two methods used above. This yields
\begin{equation}
{\cal C}_{0.1-2.4\,{\rm keV}}\approx (10\pm 2)\,\%, 
\end{equation}
with  quoted uncertainty covering the 1$\sigma$ deviations 
of all previously computed averages. A similar value 
was obtained recently by Oukbir et al. (1997).

\section{Conclusions}

We have used numerical simulations of the cluster distribution
and the de Grandi (1996) X--ray cluster luminosity function, 
to create samples of X--ray type clusters. We investigate their 
relative dipole properties, by paying particular attention to what extent
the underlying ``parent'' population cluster dipole can be recovered by the 
flux--limited subsamples.

We analysed the results based on 2 different sets of observers; random 
observers and those that have a cluster dipole structure in
reasonable agreement with that of the Abell/ACO clusters (LG-like
observers).  We found that on average the ``parent'' cluster dipole is 
underestimated by $\sim 20/30$ per cent in the former type of observers
and slightly less, $\sim 10/20$ per cent, in the latter type
(for the raw and shot-noise corrected cases respectively). 
However, the profile of the ``parent'' dipole amplitude as well as the
dipole direction is much better recovered in the LG-like observer case
($\Delta\theta \sim 30^{\circ}$ and 47$^{\circ}$ respectively for the
LG-like and random observers). We also found that the exclusion from
the flux--limited X--ray cluster samples of the local region ($\mincir
45h^{-1}\,{\rm Mpc}$), as is the case with RASS, will result into
larger underestimates of the ``parent'' dipole amplitude and larger
misalignments of the dipole directions. 

The generic loss of dipole amplitude will be important in interpreting
cosmological information coming from samples of X--ray selected clusters,
such as their relative bias factor and the inferred values of $\Omega_0$.

Finally, we attempted to estimate the contribution of these X--ray
clusters to the soft (0.1-2.4 keV) XRB for the particular case of a
flat universe with $\Omega_{\rm \Lambda}=0$. We use the standard
method for deriving the cluster volume emissivity, as well as that
using the cluster number-flux relation. The fractional contribution of
clusters to the soft ROSAT energy band is estimated to be $10 \pm 2$
per cent, using both techniques.

\vspace{1.0cm}

${\bf Note\;added\;in\;Manuscript}$: When this project was in its final stages,
two new X--ray flux--limited cluster samples appeared in the literature,
namely the X--ray Brightest Abell-type Cluster sample (XBACs; Ebeling
et al. 1996b) and Brightest Cluster Sample (BCS; Ebeling et al. 1997a
and Ebeling et al. 1997b). The XBACs sample consist of 242 Abell/ACO
X--ray clusters above the limiting flux of $5 \times 10^{-12}\,{\rm
erg}\,{\rm s^{-1}}\,{\rm cm^{-2}}$ and it is the biggest all-sky X--ray
cluster list, up till now. The BCS (largest in the northern hemisphere)
comprises 199 X--ray Abell, Zwicky as well as other purely X--ray
selected systems, being flux--limited above
$4.45 \times 10^{-12}\,{\rm erg}\,{\rm s^{-1}}\,{\rm cm^{-2}}$. Both
samples are selected at high galactic latitudes ($|\,b\,|\;\magcir
20^{\circ}$) and in the same soft energy band (0.1-2.4 keV). In both
samples the cluster identification procedure is mostly based on the
Voronoi Tesselation Percolation (VTP) algorithm (see also Allen et
al. 1992; Ebeling \& Wiedenman 1993; Crawford et al. 1995), which
not only samples nearby extended X--ray emission much more adequately
than the SASS algorithm, but also increases significantly the accuracy
with which fluxes are computed.
This, together with the fact that their depth coverage is large ($z\sim 0.2$ 
and 0.3 for XBACs and BCS respectively), renders both samples ideal
tools for large scale structure studies. These catalogues together
with the forthcoming ESO KP will definitely constitute three of the
more significant X--ray databases for future cosmological research.
We, therefore, intend to study these samples extensively along the
lines set in this work. 

\section*{\bf ACKNOWLEDGMENTS}
VK receives a PPARC research studentship. PC is a PPARC Advanced Research
Fellow. We benefited greatly from endless discussions with Luigi Guzzo
regarding some crucial observational features of the ESO KP and from
his productive criticism on a late draft of this work.
We also thank Enzo Branchini for various useful suggestions and Dimitra 
Rigopoulou and Sabrina de Grandi for invaluable information on some
parts of this work. VK also acknowledges fruitful discussions with
Jacob Sharpe and Ioannis Georgantopoulos. Special thanks to Harald
Ebeling for the long and instructive comments about several aspects of
the {\small ROSAT} surveys and cluster identification procedures and for
reading carefully an early version of the present paper. This work has
been partially supported by funds originating from the EC Human
Capital and Mobility Network (Contract Number CHRX-CT93-0129).

\newpage

\section*{\bf FIGURE CAPTIONS}

\vspace{0.4cm}
\noindent

\vspace{0.4cm}
\noindent

{\bf Figure 1.} X--ray selection function for a model ${\rm ROSAT}$-like
cluster population. 

\vspace{0.4cm}
\noindent

{\bf Figure 2.} The average, over 1500 random observers, monopole as a
function of radial distance for the three samples of clusters. 
Errorbars reflect the observer-to-observer scatter. Filled dots are
for ``parent'' (C) monopoles, open dots are for T monopoles and filled 
triangles refer to R monopoles. 

\vspace{0.4cm}
\noindent

{\bf Figure 3.} Dipole amplitude as a function of radial depth for some 
characteristic random observers. The solid line corresponds to the R, 
open circles to the T and filled dots to the ``parent'' (C) samples. The
upper 2 panels represent the most frequent  cases, lower right those
less frequent and finally the lower left are the rarest of all. 

\vspace{0.4cm}
\noindent

{\bf Figure 4.} Relative velocity fluctuations (equation 14)
for the 1500 random observers. Errorbars reflect the 
observer-to-observer 1$\sigma$ scatter and are shown only for one set
of points for clarity. Filled dots correspond to fluctuations between
the ``parent'' and T samples while open dots between the ``parent'' and R
samples.

\vspace{0.4cm}
\noindent

{\bf Figure 5.} Relative misalignment angles, $\Delta\theta$. 
random observer results are shown in panel (a) while those of the
LG--like observers are shown in panel (b). Errorbars are only shown
for one set of points, in each panel, for clarity. In both panels
filled symbols correspond to angular fluctuations between C and R
samples, whilst open symbols represent misalignment angles between C
and T populations. 

\vspace{0.4cm}
\noindent

{\bf Figure 6.} Luminosity--distance diagram for a model
{\small ROSAT}-like sample, as viewed by a typical observer. 
The solid line corresponds to the cut-off introduced due to the flux
limit imposed. The plot shows the X--ray cluster distribution of
$\sim 1300$ such objects seen by a specific observer, but the result
(shape) is independent of the observer's choice.

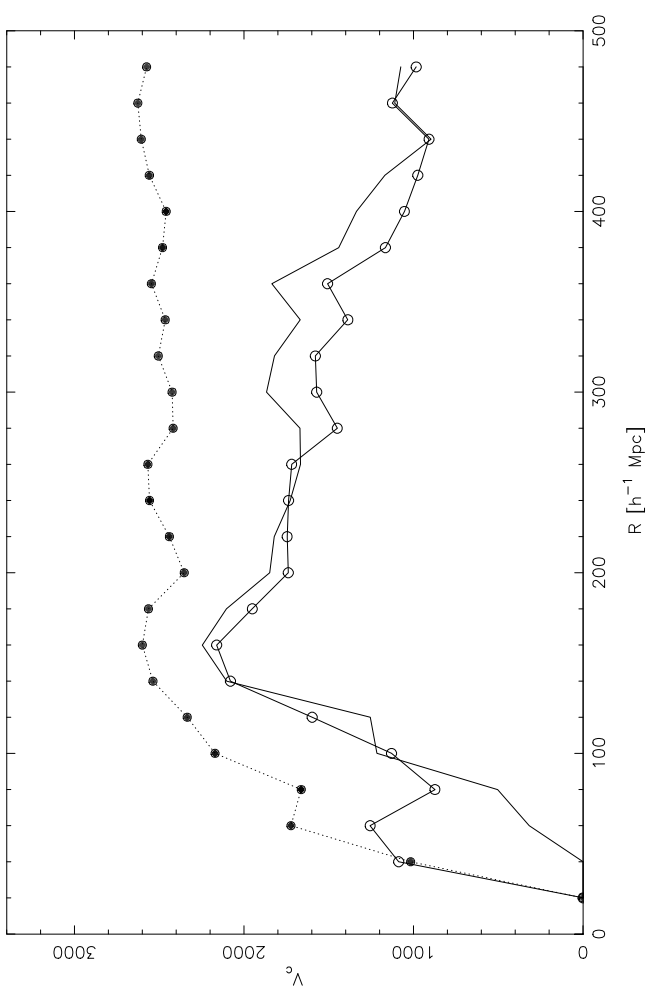
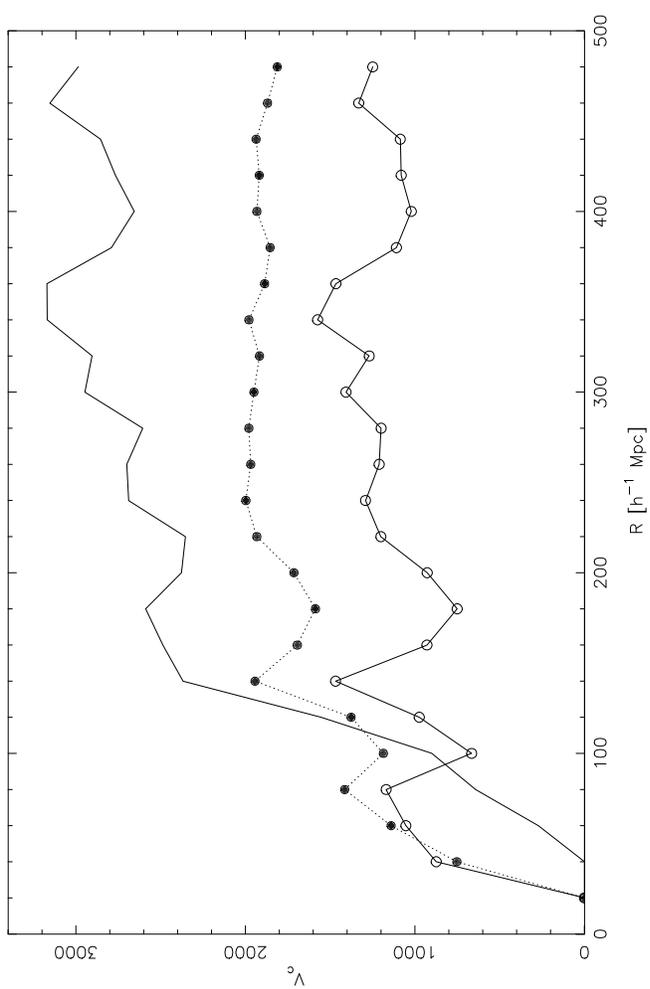
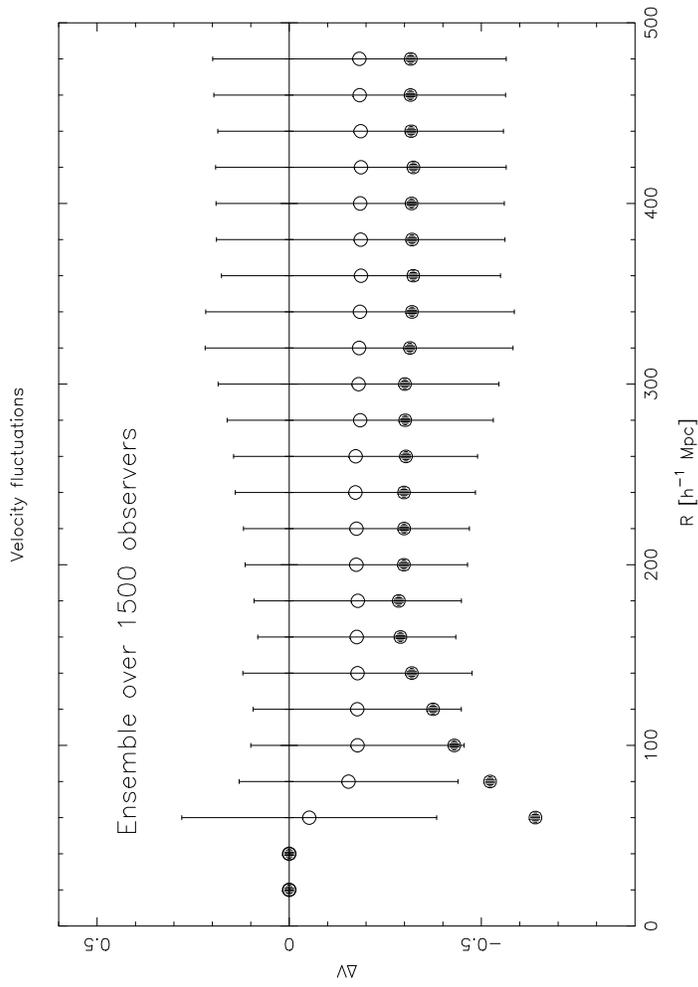
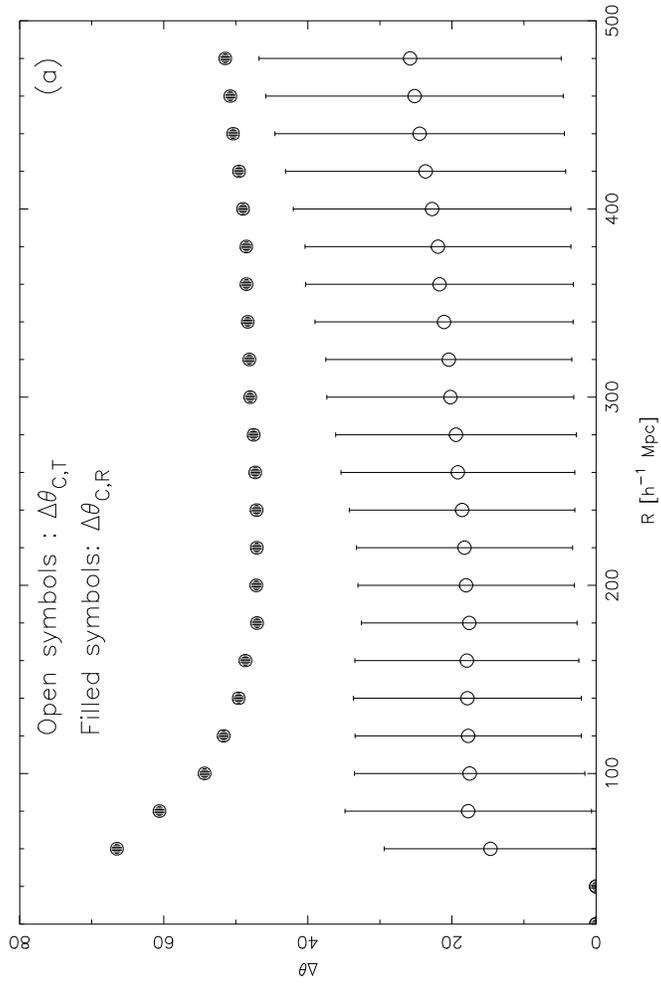
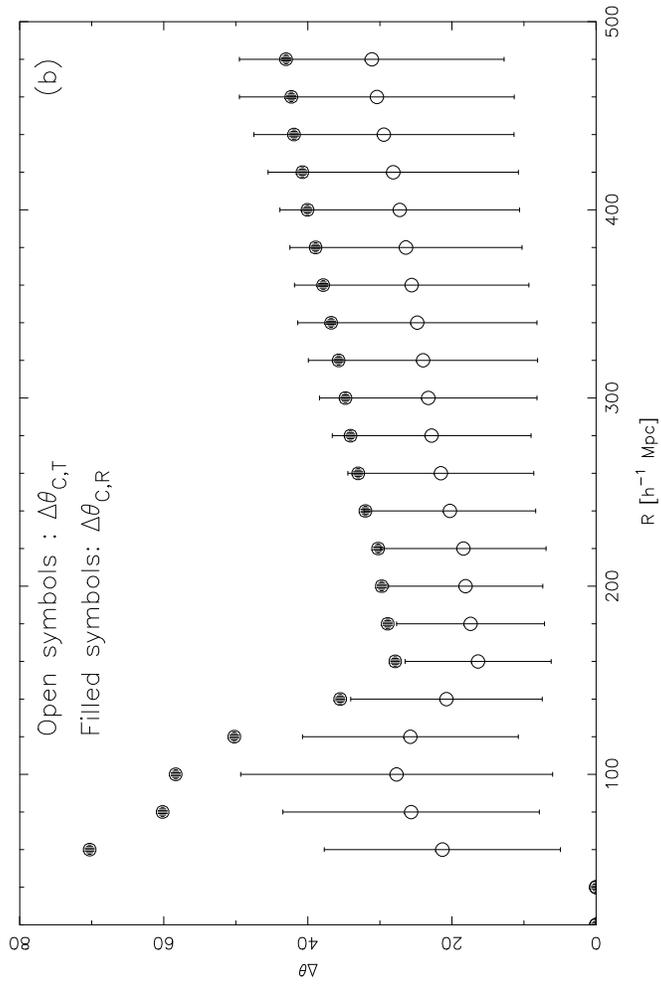
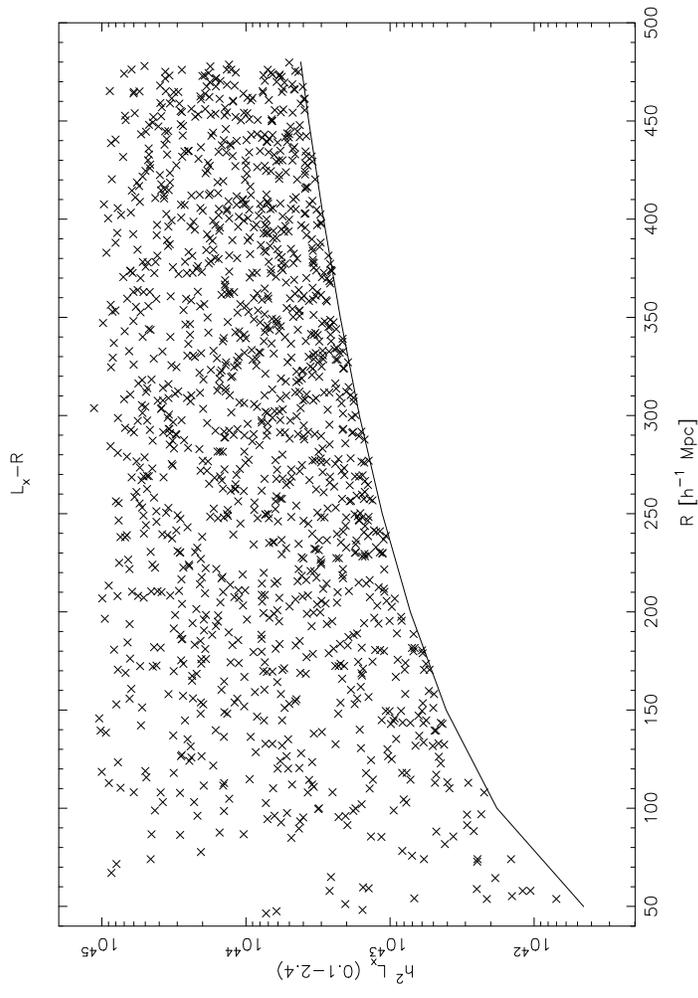
\end{document}